\documentclass{aa}  
\usepackage{graphicx}
\usepackage{txfonts}
\usepackage{color}
\usepackage[normalem]{ulem}

\usepackage{soul}
\definecolor{teal}{rgb}{0.0, 0.5, 0.5}
\definecolor{violet}{rgb}{0.56, 0.0, 1.0}
\definecolor{auburn}{rgb}{0.43, 0.21, 0.1}
\definecolor{safetybrown}{rgb}{0.7, 0.3, 0.1}
\definecolor{purple}{rgb}{0.5, 0.0, 0.5}
\begin{document} 

    %\title{Impact of solar wind turbulence on the Earth's bow shock}
    \title{Impact of solar wind turbulence on a planetary bow shock}

   \subtitle{A global 3D simulation}

   \author{E. Behar
          \inst{1, 2}
          \and
          F. Pucci\inst{3}
          \and
          C. Simon Wedlund\inst{4}
          \and
          P. Henri\inst{1, 5}
          \and
          G. Ballerini\inst{6,7}
          \and
          L. Preisser\inst{4}
          \and
          F. Califano\inst{7}
          }

   \institute{Laboratoire Lagrange, Observatoire de la Côte d'Azur, Université Côte d'Azur, CNRS, Nice, France\\
              %\email{pierre.henri@oca.eu}
         \and
             Swedish Institute of Space Physics, Kiruna, Sweden\\
        \and
            Institute for Plasma Science and Technology, National Research Council, CNR-ISTP, Bari, Italy\\
            \email{francesco.pucci@istp.cnr.it}\\
        \and
            Space Research Institute, Austrian Academy of Sciences, Graz, Austria\\
        \and
            LPC2E, CNRS, Université d'Orl\'eans, CNES, Orl\'eans, France\\
        \and
            LPP, CNRS/Sorbonne Université/Université Paris-Saclay/Observatoire de Paris/Ecole Polytechnique Institut Polytechnique de Paris, Palaiseau, France\\
        \and 
            Dipartimento di Fisica “E. Fermi”, Università di Pisa, Pisa, Italy\\
    }

   \date{Received: ... ; accepted ...}

  \abstract
   {The interaction of the solar wind plasma with a magnetized planet generates a bow-shaped supercritical shock ahead of it. Over the past decades, near-Earth spacecraft observations have provided insights into the physics of the bow shock, suggesting that solar wind intrinsic turbulence influences the bow shock dynamics.  
   On the other hand, theoretical studies, primarily based on global numerical simulations, have not yet investigated the global 3D interaction between a turbulent solar wind and a planetary magnetosphere. This paper addresses this gap for the first time by investigating the global dynamics of this interaction, providing new perspectives on the underlying physical processes.
   }
   {We examine how the turbulent nature of the solar wind influences the 3D structure and dynamics of magnetized planetary environments like those of Mercury, Earth, or magnetized Earth-like exoplanets, using the newly developed numerical model \emph{Menura}.
   }
   {We use the hybrid Particle-In-Cell model \emph{Menura} to conduct 3D simulations of the turbulent solar wind and its interaction with an Earth-like magnetized planet through global numerical simulations of the magnetosphere and its surroundings. \emph{Menura} runs in parallel on GPUs, enabling efficient and self-consistent modelling of turbulence.}
   {By comparison with a case in which the solar wind is laminar, we show that solar wind turbulence globally influences the shape and dynamics of the bow shock, the magnetosheath structures, and the ion foreshock dynamics.
   
   We show that a turbulent solar wind disrupts the coherence of foreshock fluctuations, induces large fluctuations on the quasi-perpendicular surface of the bow shock, facilitates the formation of bubble-like structures near the bow shock's nose, and modifies the properties of the magnetosheath region. None of these phenomena occur when comparing with the case in which the solar wind is laminar.
   }
   {The turbulent nature of the solar wind impacts the 3D shape and dynamics of the bow shock, magnetosheath, and ion foreshock region. This influence should be taken into account when studying solar wind--planet interactions in both observations and simulations. We discuss the relevance of our findings for current and future missions launched into the heliosphere.
   }

   \maketitle
%
%-------------------------------------------------------------------

\section{Introduction}

% Turbulent solar wind
The solar wind is a supersonic and super-Alfvénic plasma flow, mainly composed of energetic protons embedded in a large-scale magnetic field. It fills the interplanetary medium and directly interacts with planets, forming a magneto-environment around them. The main features of this environment are a supercritical collisionless bowshock, a turbulent magnetosheath and an induced elongated magnetosphere downstream of it \citep{Parks2021, Sibeck2021, Southwood2021}.  Depending on the value of $\theta_{Bn}$, defined as the angle between the local shock normal and the upstream magnetic field's direction, the bow shock can be locally classified as quasi-parallel ($\theta_{Bn}<$~45$^\circ$) or quasi-perpendicular ($\theta_{Bn}>$~45$^\circ$), with $\theta_{Bn}$ values around 45$^\circ$ defining a so-called oblique geometry \citep{jones1991, schwartz1998}. The existence of these two main shock geometries leads to different plasma kinetic dynamics around the bow-shock region \citep{burgessbook2015}. 

Solar wind-planet interaction has been extensively studied over the last decades using numerical simulations. 
%The global interaction of the solar wind and Earth-like magnetospheres has been investigated in the past by means of two-dimensional (2D) hybrid models since pioneering studies of curved collisionless shocks \citep{thomas1990}, Earth's magnetosphere models \citep{swift1995}, the interaction of interplanetary rotational discontinuity with Earth's magnetosphere \citep{lin1996} and wave behaviour \citep{lin2001} and velocity distribution functions \citep{lin2002} in the magnetosheath region. 
The global interaction of the solar wind and Earth-like magnetospheres has been investigated in the past by means of two-dimensional (2D) kinetic hybrid models, where ions are treated as individual kinetic macroparticles and electrons as a charge-neutralizing magnetohydrodynamic fluid. Pioneering hybrid modelling studies include those for curved collisionless shocks \citep{thomas1990} and for Earth's magnetosphere models \citep{swift1995}, and have, for example, focused on the interaction of an interplanetary rotational discontinuity with Earth's magnetosphere \citep{lin1996}. Later on, wave behaviour \citep{lin2001} 
and velocity distribution functions \citep{lin2002} have been studied extensively in the magnetosheath region. 
A three-dimensional (3D) geometry has been used to reproduce the basic dynamics of the magnetosphere with the existence of a turbulent magnetosheath medium, ion foreshock and waves associated with different %magnetospheric 
regions upstream of the magnetopause \citep{Kallio2003a,Kallio2004,2007GRLTravnivcek,Muller2011,Muller2012,vonAlfthan2014,Modolo2016,Jarvinen2020,Aizawa2021,Aizawa2022,Kallio2022,Teubenbacher2024}. 
Recently, global full kinetic Particle-In-Cell (PIC) simulations of the interaction between the solar wind and a magnetized planet have been performed in 2D \citep{Pengetal2015} and 3D \citep{Lavorentietal2022,Lapentaetal2022,Lavorentetal2023} to investigate the role of electron kinetics in the global interaction of the solar wind with a magnetized planet. 

All such global models always take, for the sake of simplicity, the standpoint that the solar wind plasma dynamics is laminar. 
Nevertheless, the solar wind is turbulent, with relatively large amplitude, large-scale magnetic and density fluctuations driven by continuous large-scales energy injection from the Sun. Solar wind fluctuations span a large range of spatial and temporal scales \citep{bruno2013solar,kiyani2015dissipation,Verscharen2019}. It is expected that the turbulent solar wind may influence the shock dynamics, as predicted by basic theoretical models \citep{zank2002interaction}. 

Observational studies have focused on the dynamics and turbulent nature of the solar wind and its connection to the bow shock, magnetosheath and magnetosphere dynamics \citep[see, e.g.,][and references therein]{Rakhmanova2023}. In particular, observations have shown that geomagnetic activity depends on internal magnetospheric processes and solar wind conditions \citep{damicis2020fass,guio2020fass,guio2021fass}.   
Complementary to observations, numerical studies of the interaction of solar wind turbulence with an interplanetary shock are very recent. Different with respect to global simulations, they have been performed in a `local' sense, that is, looking at a relatively small portion of the shock interaction region, not taking into account the global curved nature of a planetary shock and using one wall of the simulation as a fully reflective boundary. These local hybrid PIC simulations have shown that turbulent fluctuations in the upstream region enhance particle acceleration at the shock front, leading to a diffusive spread of the particles in velocity space \citep{trotta2021phase}. This result has been supported by observations of an increase in the magnetic helicity downstream of the shock as turbulent structures are compressed while transmitted across the quasi-perpendicular shock \citep{guo2021shock, trotta2022transmission}. 
Local hybrid PIC simulations have also been used to study the interaction of multiple current sheets with a shock wave, discussing the implication of such interaction on particle acceleration in the downstream shock region \citep{nakanotani2021interaction}. Further hybrid PIC simulations have confirmed the role of upstream turbulence as a scattering agent to promote diffusive shock acceleration \citep{nakanotani2022turbulence}.
More recently, by coupling turbulent MHD fields and local quasi-perpendicular hybrid kinetic 3D simulations, \cite{Trotta2023}
showed that turbulence increases fluctuations at the shock interface and the isotropization of the magnetic field spectra in the downstream region close to the bow shock.

However, none of the above studies have investigated the global response of a magnetised planet's magnetosphere to solar wind turbulence. The numerical model that we use in this study, namely \emph{Menura}, has been specifically designed for this purpose. \emph{Menura} can self-consistently model a fully-developed turbulent solar wind interacting with a planet \citep{behar2022angeo} and was recently used by \cite{behar2023aa} to show in 2D that the turbulence of the solar wind significantly modifies the dynamics of the induced magnetosphere of comets.

Here, we present the results of the first 3D hybrid simulation of the interaction between a turbulent solar wind and the magnetosphere of a magnetised planet with a size approaching that of the Earth. For the first time, we show how the turbulent nature of the solar wind affects the global shape and dynamic of the bow shock, the fluctuations in the magnetosheath, and the ion foreshock region. 
   
The paper is organized as follows.  In Section \ref{model}, we describe the model and the parameters of the simulations conducted with \emph{Menura}. Section \ref{impact} discusses the shape of the bow shock and its dynamics. We focused specifically on the ion foreshock and the structures locally created by the upstream turbulence as examples of kinetic effects captured by Menura on both quasi-parallel and quasi-perpendicular sides of the bow shock. 
We conclude and discuss the future perspectives that open up with this study in Sect.~\ref{conclusion}.

\section{The model}\label{model}

We use the 3D hybrid kinetic particle-in-cell (PIC) model \emph{Menura} to simulate the interaction of a turbulent solar wind with a planetary magnetosphere. A detailed code description is available in  \cite{behar2022angeo}, and an example of application in a reduced, so-called 2.5D geometry is described in \cite{behar2023aa}. In this work, we use the 3D version of the code. \emph{Menura} is a hybrid PIC model that provides a kinetic description of ion dynamics and employs a generalized Ohm's law coupled to a polytropic closure for the massless electrons. \emph{Menura} uses the CAM scheme (or Current Advanced Method) \citep{matthews1994current}, used routinely in PIC hybrid codes as well more recently in hybrid Eulerian Vlasov codes \citep{valentini2007jcp}. 
    
Our study here is conducted in two successive steps. 
First, we perform a 3D simulation of solar wind decaying turbulence using periodic boundary conditions. This simulation follows the solar wind evolution until a quasi-stationary state is achieved and the turbulence is fully developed. 
Second, we use the last iteration step of the turbulent decay simulation as the initial condition of a new run in which we simulate the interaction of this turbulent solar wind with a compact magnetized planet (such as the Earth or Mercury).
More details on treating boundary conditions for this second step are described in \citet{behar2022angeo} and \citet{behar2023aa}. 
Additionally, a reference simulation is performed using laminar solar wind conditions to properly assess the effects of solar wind turbulence on the interaction with the planetary obstacle.
    
In both steps, the solar wind and planetary plasma dynamics equations are solved in the solar wind reference frame. 
Unlike the object-centred reference frame used in other models with similar scientific purposes \citep{vonAlfthan2014,Grandin2023,Karimabadi2006}, solving equations in the solar wind reference frame enables the introduction of a solar wind flow-aligned magnetic field that varies in time. This condition is necessary to inject a well-defined, fully developed, self-consistently generated turbulent flow that includes, for example, magnetic vortices. 
    
In the following, we describe the initial conditions and parameters of these two successive simulations, which we named Sim 1 and Sim 2.1, and the reference laminar run, which we named Sim 2.2.    Table~\ref{tab:turbulenceInputs} summarises the simulations' input parameters. 

\begin{table*}
    \centering
    \begin{tabular}{lcrl}
         Variable       & Unit      & Value & Description\\
         \hline
         Normalisation  &           & & \\
         Density & $n_0$          & & Initial solar wind density\\
         Magnetic field & $B_0$          & & Module of initial solar wind magnetic field\\
         Time & $\Omega_\text{ci}^{-1}$ & $m_p c/eB_0$ & Inverse of proton gyrofrequency\\
         Speed & $C_A$      & $B_0/\sqrt{4\pi n_0 m_p}$& Alfvén speed\\
         Length & $d_i$          & $C_A/\Omega_\text{ci}$& Proton inertial length\\
         Pressure & $P_0$ & $n_0 m_p C_A^2$ & Normalizing pressure\\
         %$\Omega_\text{ci}$ & s$^{-1}$ & 0.287 & Proton gyrofrequency\\
         %$d_i$          & km      %  & 131.5 & Proton inertial length\\
         %$d_e$          & km      %  & 3.1 & Electron inertial length\\
         %$B_0$          & nT      %  & 3.0 & Module of initial solar wind magnetic field\\
         %$n_0$          & %cm$^{-3}$ & 3.0 & Initial solar wind density\\
         %$C_A$          & %km\,s$^{-1}$ & 37.8 & Alfvén speed\\
         %$P_0 = B_0^2/\mu_0$ & %nPa       & $7.16\times10^{-12}$ & Normalising ion pressure\\
         \hline
         Solar wind parameters   & & &\\ 
         $n_\text{sw}$  &   $n_0$  & $1.0$ & Solar wind proton density \\
         $\mathbf{B}_\text{sw}$ & $B_0$    & $\left(\frac{1}{\sqrt{2}}, \frac{1}{\sqrt{2}}, 0\right)$ & Solar wind magnetic field vector\\
         $\mathbf{U_\text{sw}}$ & $C_A$    & $\left(-10, 0, 0\right)$ & Solar wind velocity vector,  planet's reference frame\\
         $T_i/T_e$        &    -     & $1.0$ & Ratio of proton temperature to electron temperature\\
         $\beta_\text{sw}=\beta_i+\beta_e$    &  -       & 1.0 & Solar wind plasma beta\\
         $\gamma_e$    & -       & 1.0 & Electron adiabatic index\\
         %$C_A$  & km\,s$^{-1}$ & 37.8\\
         %$C_\text{ms}$ & km\,s$^{-1}$ & 65.4 & Magnetosonic speed\\
         $C_\text{s}$ & $C_A$ & 1.4 & Solar wind sound speed\\
         $C_\text{ms}$ & $C_A$ & 1.7 & Solar wind magnetosonic speed\\
         $M_\text{A}$ & - & 10 & Bow shock Mach number\\
         %$\eta_\text{hyp}$&           & 0.01\\
         \hline
         Turbulence    &       & &\\
         $\delta B_0/B_0$ & -     & 0.54 & Initial magnetic field fluctuations\\
         $\delta \varv_0/C_A$ & -     & 0.54 & Initial velocity fluctuations\\
         \hline
         Magnetized planet  & & \\
         $\left(X_\text{P},Y_\text{P},Z_\text{P}\right)$ & $d_i$ & $\left(\frac{3L_{box}}{8},\frac{L_{box}}{2},\frac{L_{box}}{2}\right)$ & Position of the planet's centre in simulation box \\
         $\tau_\text{dip}$ &   -    & $\left(0, -1, 0\right)$ & Dipole moment direction\\
         $D_\text{mp}$ &   $d_i$    & $200$ & Distance to the magnetopause from the planet's centre\\
         \hline
         Grid and numerics       &           & &\\
         $\Delta X=\Delta Y=\Delta Z$     & $d_i$     & 5.0 & Grid resolution\\
         $\Delta t$     & $\Omega_\text{ci}^{-1}$ & 0.5 & Time resolution\\
         $L_{box}=L_X=L_Y=L_Z$& $d_i$     & 2000 & Box size in each spatial direction\\
         $N_\text{pcc}$ & -         & 600 & Number of particles per cell\\
         $\eta_\text{hyp}$&   -     & 0.01 & Numerical hyper-resistivity\\
         \hline
         Simulation names       &           & &\\
         Sim 1       & & & Decaying solar wind turbulence\\
         Sim 2.1     & & & Turbulent solar wind vs. planet\\
         Sim 2.2     & & & Laminar solar wind vs. planet\\
         \hline
    \end{tabular}
    \caption{Input parameters of the simulations performed: code normalizations, solar wind parameters for all runs, initial amplitude of turbulence in the decaying run,
    characteristics of the magnetized planet, grid and time resolution for all runs. $m_p$ is the proton mass, $c$ the speed of light in vacuum and $e$ the elementary charge. Simulation parameters are typical of the solar wind \citep{Owens2023}, with $n_0$ and $B_0$ the initial solar wind density and magnetic field magnitude.}
    
    \label{tab:turbulenceInputs}
\end{table*}

\subsection{Decaying simulation of solar wind turbulence (Sim 1)}
\label{sec:simulation1}

In this kinetic hybrid simulation, the solar wind consists of one ion species, i.e. protons, and massless neutralizing electrons. 
%--- Box
The simulation domain is a Cartesian box of equal size $L_{box} = L_X = L_Y = L_Z = 2000\ d_i$ in the three spatial directions, discretized in $400$ cells in each direction with a spatial resolution of $\Delta x = 5 d_i$, with $d_i$ the solar wind proton inertial length. We populate each cell with $600$ macroparticles to ensure a statistically satisfactory representation of the ion distribution function.
%--- Time step
The simulation's time step is $\Delta t= 0.5 \ \Omega_{ci}^{-1}$, with $\Omega_{ci}$ the solar wind proton gyrofrequency. 
%--- Comment 
Consequently, these simulation parameters are such that ion scales are poorly resolved spatially and temporally. Such a resolution is imposed by computational constraints; however, in this study, we do not specifically focus on the dynamics at the ion and sub-ion scale but rather on phenomena just below the smallest MHD scales, approaching the ion kinetic scales, while enabling us to describe some kinetic features such as a supercritical bow shock and the associated reflected ions in the foreshock. 
%--- Initial conditions
The initial equilibrium condition is made of a solar wind plasma with homogeneous density and temperature, permeated by a homogeneous, oblique (to the solar wind flow) mean magnetic field $B_0$ (see Table~\ref{tab:turbulenceInputs}). At equilibrium, the ratio of ion kinetic and magnetic pressures is $\beta_i=0.5$ and the ion-to-electron temperature ratio is $T_i/T_e=1$, resulting in $\beta=\beta_i+\beta_e=1$.
We impose an isothermal closure on electrons, corresponding to an adiabatic index $\gamma_e=1$. 
We perturb this equilibrium with magnetic and velocity fluctuations at large scales. The initial velocity fluctuations are incompressible following $\nabla\cdot{\bf v}=0$. 
The initial perturbation is made of sinusoidal fluctuations with a polarization orthogonal to both the mean field and the wavevector ${\bf k}$. The wavevectors are directed along the three Cartesian directions and all wavevectors within the range $[k_{min}, k_{max}] = [2 \pi /L_{box}, 5 \cdot 2 \pi/L_{box}]$ are populated. The phases are random and different for the velocity and magnetic fluctuations. 

\begin{figure*}
\centering
\includegraphics[width=\linewidth]{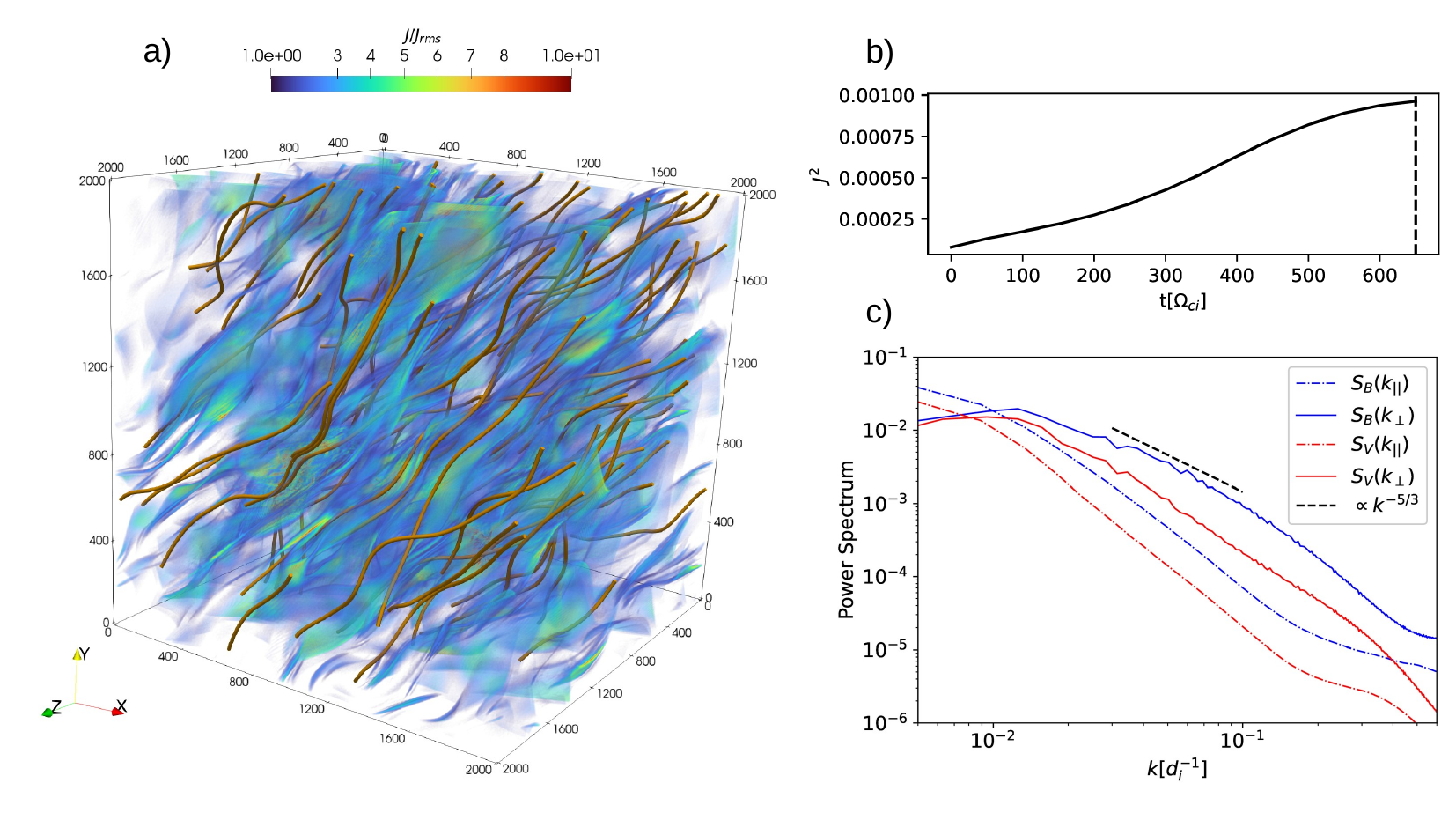}
    \caption{Characteristics of the decaying simulation of solar wind turbulence (Sim 1). a) Current density normalized to its root mean square $J_\text{rms}$ (colour map) and magnetic field lines (orange) in the full 3D plasma box. b) Box-averaged square current density $J^2$ as a function of time. The vertical dashed line marks the time of the snapshot ($t \simeq 650\,\Omega_\text{ci}^{-1}$) used to initialize the solar wind turbulence in Sim 2.1. c) Power spectrum of magnetic and velocity field for parallel and perpendicular wavevectors.}  
\label{fig:spectra_decay}
\end{figure*}

The magnetic field lines and the total charge current $|J|$ in the simulation box at the end of the decaying turbulence simulation are shown in Fig.\ \ref{fig:spectra_decay}a. 
The anisotropy in the magnetic field fluctuations is evident from the elongated shape of the current structures aligned parallel to the mean solar wind magnetic field.

The time evolution of the charge current fluctuations $J_\text{rms}$, defined as its root mean square (RMS), is shown in Fig.\,\ref{fig:spectra_decay}b. The vertical dashed line indicates the time the decaying turbulence simulation is fully developed so that it can be injected later in the magnetised planet simulation. We identify it with the time when the RMS current saturates.  
At the end of this first simulation, the RMS value of the final perturbation is $\delta B/B_0=0.45$ and $\delta \varv/C_A=0.33$, where $\delta B$ and $\delta \varv$ are the magnetic and velocity RMS values, $B_0$ is the background magnetic field and $C_A$ is the background Alfvén speed in normalised units. 
        
We have computed the parallel and perpendicular (to the mean solar wind magnetic field direction) spectra of magnetic and velocity fluctuations, shown in Fig.\,\ref{fig:spectra_decay}c. The magnetic field follows a power-law trend with a spectral slope consistent with the expected Kolmogorov decay of $-5/3$. At smaller scales, closer to one $d_i$, the spectral trend changes under the effect of both the numerical dissipation (hyper-resistivity) and dispersive and kinetic ion physics \citep{matteini2016}. 

The electric field and magnetic field as well as the plasma distribution function from the decaying turbulence simulation at time $t \simeq 650\,\Omega_\text{ci}^{-1}$ are used to initialize the second simulation of our model (Sim 2.1).
    
\subsection{Interaction between a magnetized planet and a solar wind turbulent dynamics (Sim 2.1)}\label{sec:simulation2}

In this second simulation, we model the interaction between a solar wind with fully developed turbulent dynamics, resulting from %CPthe decaying turbulence run of Sect.\,\ref{sec:simulation1}
Sim 1, and a magnetized planet. 
The magnetised planet is modelled as a perfectly absorbing body on which entering ions are removed from the simulation, together with a permanent magnetic dipole, taken as an external magnetic field. 

Since the computation is performed in the solar wind frame, the planet is moving in the simulation domain %solar wind 
at the opposite of the solar wind speed. 
To maintain the planet at a fixed position in the simulation domain, we continuously shift the domain sideways (in the $+X$ direction). Consistently, the dipole field is recalculated each time the simulation box moves. In fact, our choice of reference frame requires adding an additional term in Faraday's law corresponding to a Lorentz transformation. To our knowledge, this is the first time a non-fixed reference frame has been used for this type of application.
In the simulation box, the planet's centre is kept at coordinates $\left(X_\text{P},Y_\text{P},Z_\text{P}\right) = \left(3L_\text{box}/8,L_\text{box}/2,L_\text{box}/2\right) = \left(750,1000,1000\right)\,d_i$, with $L_\text{box}$ the size of the box in any direction in units of $d_i$.

The dipole value is chosen by defining the value $D_p$ as the position of the nose of the magnetopause normalized to the ion's inertial length. This parameter has proven to be an effective method for characterizing the magnetospheric structure as a function of dipole strength \citep{omidi_dipolar_2004, Karimabadi2014}. In this work, to reduce the computational effort, we use $D_p = 200\,d_i$, a smaller value with respect to the real value at Earth ($D_{p,\text{Earth}}\sim 640\,d_i$). As pointed out in \cite{omidi_dipolar_2004}, simulations with $D_p$ greater than $\sim 20$, one order of magnitude smaller than the one we used, have Earth-like characteristics both on the dayside and in the magnetotail. The smaller size of the magnetosphere and magnetosheath reduces the transit time of the plasma inside the magnetosheath by a factor~$\sim 3$ with respect to the Earth. This may affect the development of waves in the region, such as wave modes with a relatively low growth rate, as they may not have time to develop before reaching the magnetopause.
However, this work is the first step in studying how turbulent solar wind globally affects the different large-scale frontiers in a planetary magneto-environment. %[Etienne: great paragraph!]  

%The region the closest to the body is notoriously difficult to simulate since magnetic dipole field values follow a $r^{-3}$ profile, resulting close to the centre of the dipole in unfeasible constraints on the time step $\Delta t$ with regards to the grid spacing $\Delta X$. We have also observed that it is mostly the transition period, from a non-existent to an established magnetosphere, which poses the greatest stability issues during the runs, while once the magnetosphere has reached an average steady state, the system is much more stable. During our test runs, we have noticed that applying a smoothing, which eliminates high wavenumber fluctuations, on the magnetic field at each time step solves the instability issues that may arise in this transition phase. However, in the production runs presented in this paper, such a smoothing procedure has not been used since the runs remained stable also during the transition phase.    

\begin{figure*}
    \centering
    \includegraphics[width=.75\linewidth]{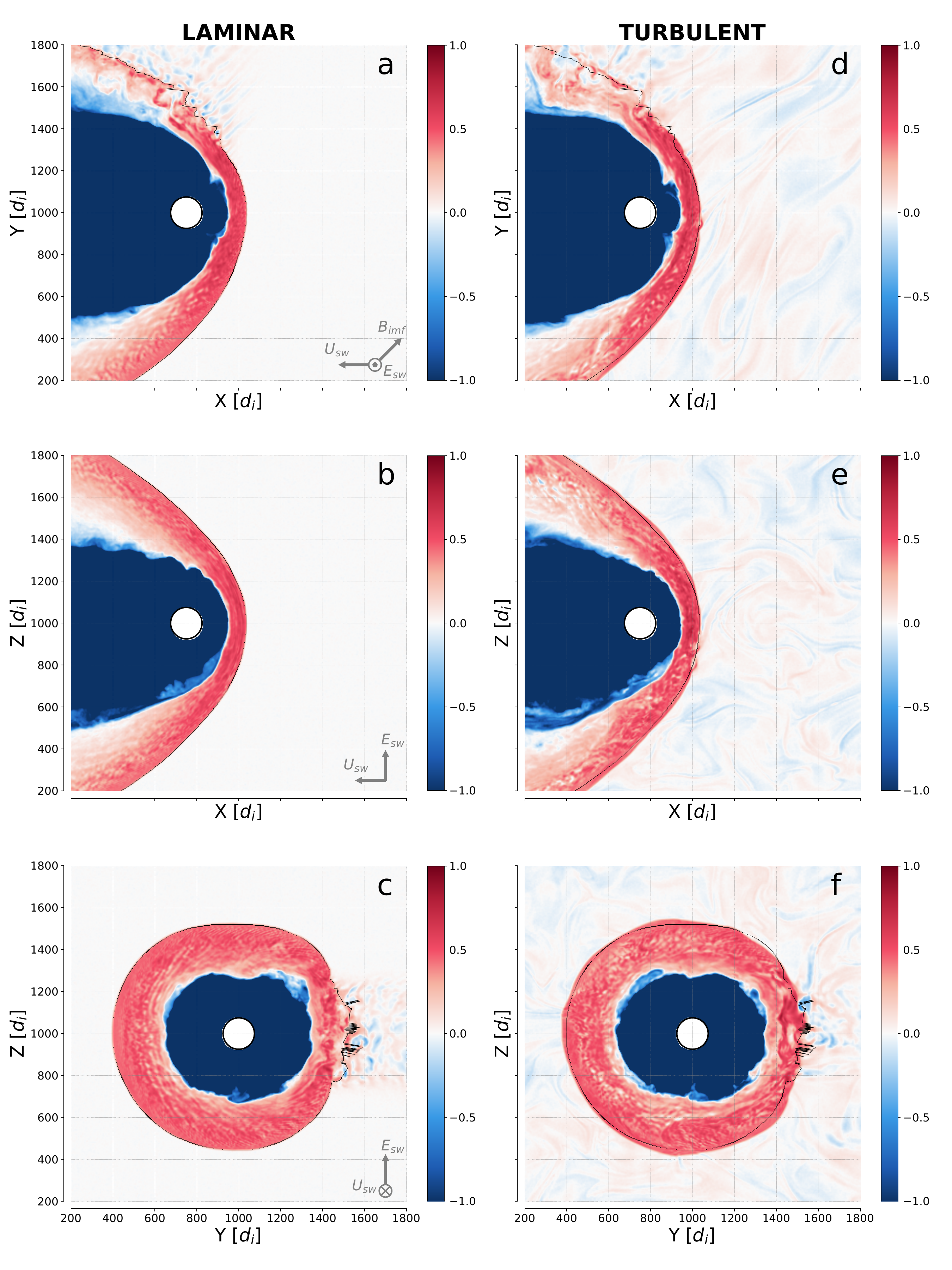}
    \caption{Comparison of ion density in logarithmic scale between laminar (a-c) and turbulent (d-f) simulations at $t=250\,\Omega_\text{ci}^{-1}$ for the same 2D planes intersecting the planet's centre. The initial solar wind magnetic field $\mathbf{B}_\text{sw}$ is contained in the $X$--$Y$ plane (in a $45^\circ$ angle). In the planet's reference frame, the $+Z$ direction contains the solar wind convection electric field $\mathbf{E}_\text{sw}$, whereas the solar wind bulk velocity $\mathbf{U}_\text{sw}$ is along $-X$, and given as indicators. The position of the bow shock based on a density threshold for the laminar simulation (left column) is shown for comparison as a black contour line for both simulations.
    }
    \label{fig:comp_dens}
\end{figure*}

\begin{figure*}
    \centering
    \includegraphics[width=.75\linewidth]{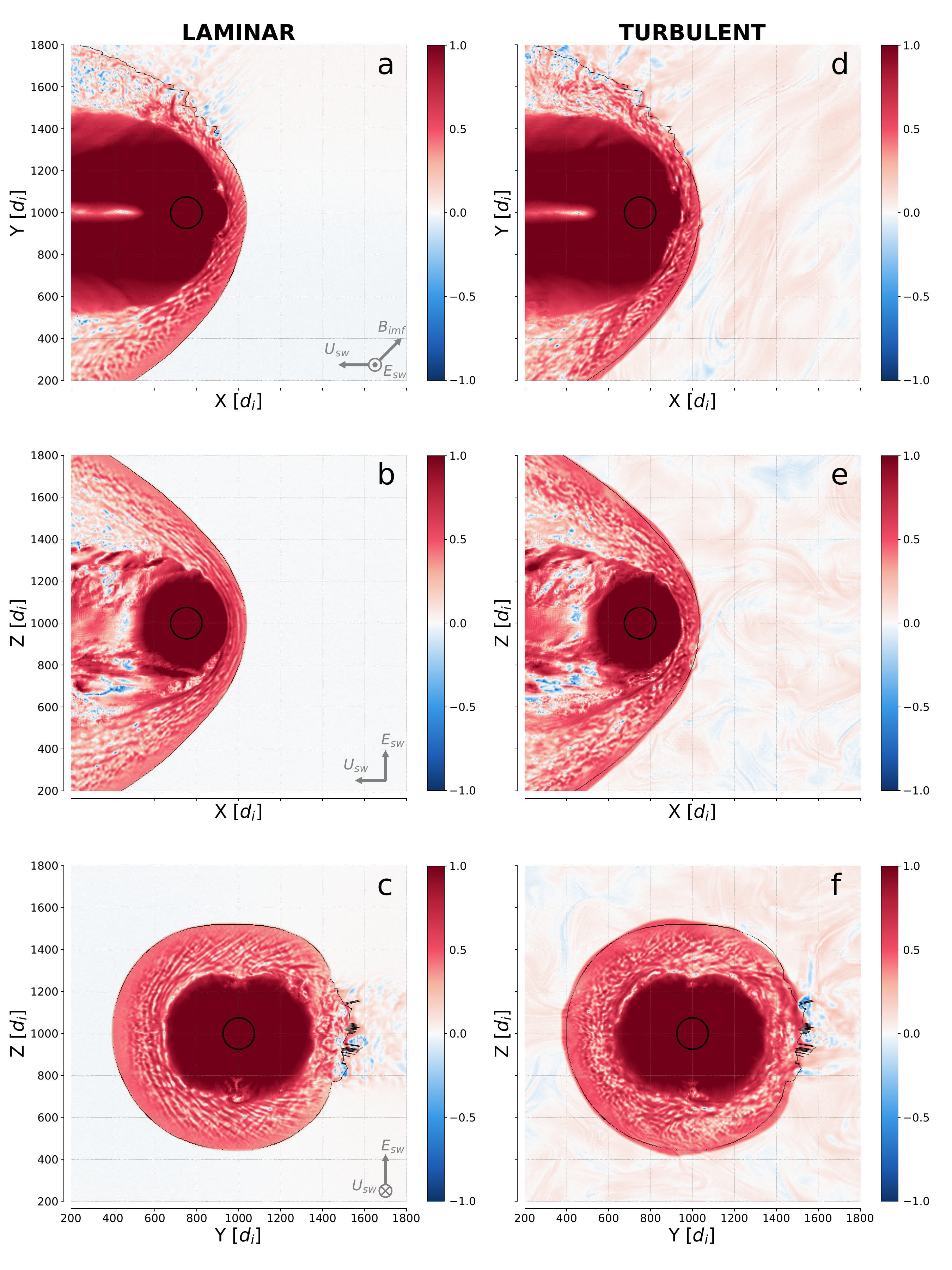}
    \caption{Comparison of the magnetic field amplitude in logarithm scale between laminar and turbulent simulations at $t=250\,\Omega_\text{ci}^{-1}$. Same format as in Fig.\,\ref{fig:comp_dens}.}
    \label{fig:comp_B}
\end{figure*}

\begin{figure*}
    \centering
    \includegraphics[width=.75\linewidth]{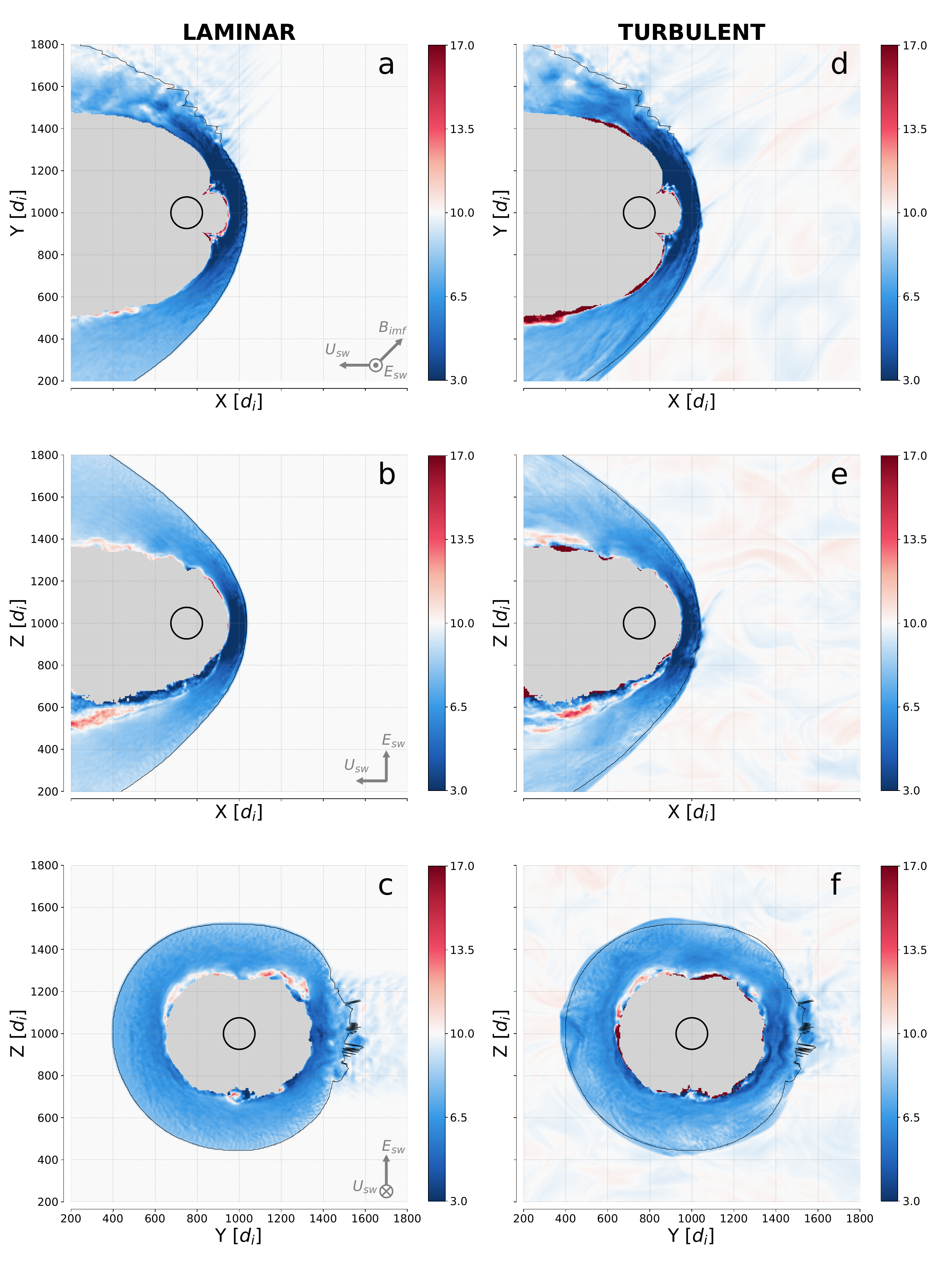}
    \caption{Comparison of the proton bulk speed $U_p$ between laminar (left column) and turbulent solar wind simulations (right column) at $t=250\,\Omega_\text{ci}^{-1}$. Note that the colour scale is linear for clarity, unlike in Figs.~\ref{fig:comp_dens}~and~\ref{fig:comp_B}. Same format as in Fig.\ \ref{fig:comp_dens}, with colour bar in units of Alfvén speed $C_A$ and white the input solar wind speed ($U_\text{sw} = 10\,C_A$). Note that the bulk speeds are expressed in the planet's reference frame here. Grey regions are those where the plasma density is smaller than $0.1 n_0$.}
    \label{fig:comp_bulkspeed}
\end{figure*}

\subsection{Interaction between a magnetized planet and a solar wind laminar dynamics (Sim 2.2)}\label{sec:sim_laminar}

To properly assess the impact of the turbulent nature of the solar wind on a magnetosphere, it is necessary to compare its effect to that of an upstream solar wind that would be laminar.
For this purpose, we run a third reference simulation in which the planet interacts with a laminar solar wind. 
In this case, the planet moves into a homogeneous solar wind with plasma density and temperature equal to those chosen as the initial condition of Sim 1. 
The solar wind magnetic field is also homogeneous and equal to $\mathbf{B}_\text{sw}$.
All other simulation parameters, including planet parameters and spatial and temporal resolution, are identical to those of Sim 2.1. 
    
\section{Impact of a turbulent solar wind on a planetary bow shock}\label{impact}

In the following, we compare the \emph{turbulent} (Sim 2.1) and  \emph{laminar} (Sim 2.2) simulations to highlight the effects that the turbulent nature of the solar wind has on the magneto-environment of a planet.
To compare the structure and dynamics of the solar wind, shock, and magnetosheath in the two simulations, we present maps of relevant quantities in three perpendicular planes intersecting the centre of the planet, located at coordinates $(X_\text{P}, Y_\text{P}, Z_\text{P})$. 
    
Figures~\ref{fig:comp_dens}, \ref{fig:comp_B} and \ref{fig:comp_bulkspeed} present global maps of the plasma density, magnetic field magnitude and proton bulk speed, respectively: the left column (panels a, b, c) shows the laminar solar wind case results, whereas the right column (panels d, e, f) shows the turbulent solar wind case at the same simulation time $t=250\,\Omega_\text{ci}^{-1}$. Density is normalised to the solar wind proton density, the magnetic field to the solar wind magnetic field, and proton bulk speeds to the Alfvén speed (see Table~\ref{tab:turbulenceInputs}).
For both turbulent and laminar simulations, the bow shock, magnetosheath, and magnetopause regions can be clearly identified, with the quasi-parallel ($Y>1400\,d_i$) and quasi-perpendicular ($Y<800\,d_i$) sides of the shock having shapes and extents in good qualitative agreement with other global simulations \citep{Turc2023}. Closer to the planet, regions where ions are seen flowing within the magnetosphere of the planet take the shape of highly structured cones in 3D (see Fig.\,\ref{fig:comp_dens}a in the $X$--$Y$ plane at $Z=Z_\text{P}$, with $Z_\text{P}$ the position of the planet's centre), closely mimicking the Earth's plasma cusps. These ``cusps'' appear relatively less defined in the turbulent solar wind case, owing to the less homogeneous magnetosheath (Fig.\,\ref{fig:comp_dens}b). 
%To complement these 2D cuts and give an impression of the 3D variability of the system under the two solar wind dynamics configurations, we present also in Appendix~\ref{appendix} a 3D tomographic view of the ion density distribution around the planet along the Sun-planet line, from the solar wind (to the right) to the flank magnetosheath (to the left). 
%Comparing the two simulations, we observe that the bow shock, the magnetosheath and the ion foreshock (in the quasi-parallel shock region) show significant changes in overall shape, extent and characteristics, as seen in all three maps of density (Fig.\,\ref{fig:comp_dens}), magnetic field (Fig.\,\ref{fig:comp_B}) and proton bulk speed (Fig.\,\ref{fig:comp_bulkspeed}). 
The following sections provide a detailed description of how the turbulent nature of the solar wind shapes these boundaries and regions. 

%==============================
\subsection{Shape of the bow shock} \label{shape}

To facilitate comparisons of the shape of the bow shock in the absence or presence of turbulence in the solar wind, we built a simple proxy of the 3D position of the shock surface saved at high temporal cadence during a numerical run.
This proxy is defined using the plasma density: for each $(Y,Z)$ coordinate, the position of the bow shock is estimated to be the first position along the $X$ direction at which the density jumps above a value of $10^{1/4}$, i.e. about $1.8$ times the solar wind background value, which is chosen as an intermediate value between the upstream solar wind and the downstream magnetosheath plasma.
    
The position of the shock in the laminar solar wind case (Sim~2.2) is shown in Fig.~\ref{fig:comp_dens} by the thin solid black line within each plane. The same line is superimposed onto the results of the turbulent case (Sim 2.1) as a baseline for comparison between the laminar and turbulent solar wind-planet simulations. Moreover, the same bow shock position in the laminar case is superimposed onto the magnetic field maps (Fig.\,\ref{fig:comp_B}), showing how well it captures the sharp transition between the upstream solar wind magnetic field (in white) and the compressed downstream magnetic field and denser magnetosheath (in red). Similarly, this sharp transition is seen on the proton bulk speed maps (Fig.\,\ref{fig:comp_bulkspeed}) where the solar wind (in white) is abruptly slowed down to subsonic speeds at and downstream of the shock (blue hues).
 
When observing the quasi-perpendicular region of the bow shock, the density maps in (Fig.~\ref{fig:comp_dens}a-f) show that the shock surface is inflated or deflated with respect to the laminar case when the impinging initial solar wind is turbulent. This feature is also confirmed by the magnetic field (Fig.\,\ref{fig:comp_B}) and the proton bulk speed maps (Fig.\,\ref{fig:comp_bulkspeed}). These fluctuations of the quasi-perpendicular bow shock surface result from local inhomogeneities in the solar wind bulk dynamic pressure, which stem from the turbulent nature of the initial solar wind condition in Sim 2.1: turbulence causes certain regions to experience higher or lower values of solar wind dynamic pressure compared to the laminar case (Sim 2.2).

The difference between the bow shock's location in the two runs is most pronounced for the quasi-parallel shock. 
In the quasi-parallel shock region, the shock surface proxy does vary widely, as seen for $Y\gtrsim 1300\,d_i$ in Fig.~\ref{fig:comp_dens}a and at $Y\sim 1500\,d_i$ in the corresponding perpendicular plane in Fig.~\ref{fig:comp_dens}c, laminar case, but can capture the overall shape of the shock in this highly variable region. That said, the sharp and well-defined transition between the upstream and quasi-parallel downstream domains in the laminar case is mostly lost in the turbulent case due to fluctuations in the solar wind that locally change the magnetic field orientation with respect to the shock normal. In this way, the density variation proxy used for the laminar case cannot capture the highly variable quasi-parallel shock interface in the turbulent case. However, such a proxy remains useful to highlight how far turbulence changes the quasi-parallel shock location and shape.
%In the quasi-parallel shock region, the sharp transition between the upstream and downstream domains in the laminar case is lost in the turbulent case due to the presence of fluctuations in the solar wind that locally change the magnetic field orientation with respect to the shock normal.
%As a result, the quasi-parallel shock surface proxy varies widely, as seen for $Y\gtrsim 1300\,d_i$ in Fig.~\ref{fig:comp_dens}a and at $Y\sim 1500\,d_i$ in the corresponding perpendicular plane in Fig.~\ref{fig:comp_dens}c. The density variation proxy used for the laminar case is unable to capture the highly variable quasi-parallel shock interface in the turbulent case. However, such a proxy is useful to highlight how far turbulence changes the quasi-parallel shock location and shape. 

In the density maps in Fig.\,\ref{fig:comp_dens}, we observe that the compression downstream of the quasi-parallel shock is less pronounced in the turbulent case  (Fig.\,\ref{fig:comp_dens}d) and the structure of the bow shock is significantly more perturbed than in the laminar case (Fig.\,\ref{fig:comp_dens}c,f). For $Y>1600\,d_i$, it becomes difficult to identify the exact location of the quasi-parallel shock boundary (Fig.\,\ref{fig:comp_dens}d).

\begin{figure*}
\centering
\includegraphics[width=\linewidth]{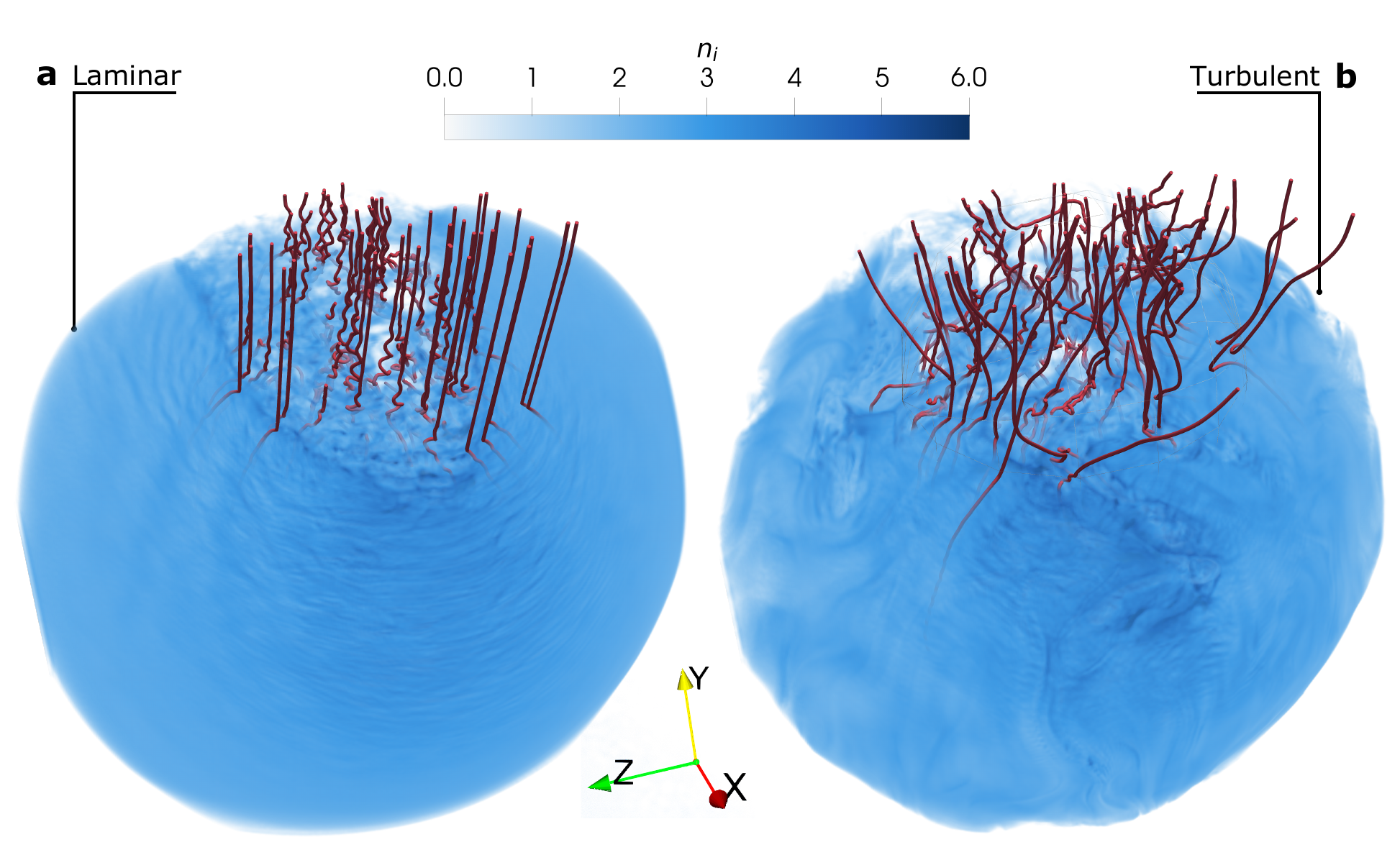}
\caption{3D rendering of the bow shock for the a) laminar and b) turbulent solar wind cases. Ion density is represented in blue hues. A threshold density $n_{th}=10^{1/4}n_0$ is applied, such that all regions in which $n_i<n_{th}$ are made transparent. A linear transparency profile is applied from $n_i = n_{th}$ to $n_i = 6 n_0$, so low-density regions are more transparent than high-density ones. Upstream magnetic field lines crossing the ion foreshock regions are drawn in red. 
} 
\label{fig:B_topology_qpara}
\end{figure*}

These differences are clearly shown in Fig.\,\ref{fig:B_topology_qpara} where the bow shock is visualized in 3D, setting a transparency threshold of $n_{th}=10^{1/4}n_0$ on the plasma density. While in the laminar case (Fig.\,\ref{fig:B_topology_qpara}a), the shock boundary is mainly smooth over all the corresponding quasi-perpendicular surface, for the turbulent simulation (Fig.\,\ref{fig:B_topology_qpara}b) large fluctuations are presented over all the bow shock. The quasi-parallel region is more easily identified in the laminar case, where the fluctuations delimit a clear area around the north pole region that corresponds to the foot points from where magnetic field lines (in red) parallel to the local shock normal are emerging. In contrast, the corresponding quasi-parallel region is not well delimited for the turbulent case, and the magnetic field lines do not appear aligned as they are in the laminar case.
This feature affects the dynamics of the ion foreshock, as discussed in more detail in Sect.\,\ref{sec:ion_foreshock}.

%==============================
\subsection{Dynamics of the bow shock} \label{dynamics}

\begin{figure*}
\centering
\includegraphics[width=.65\linewidth]{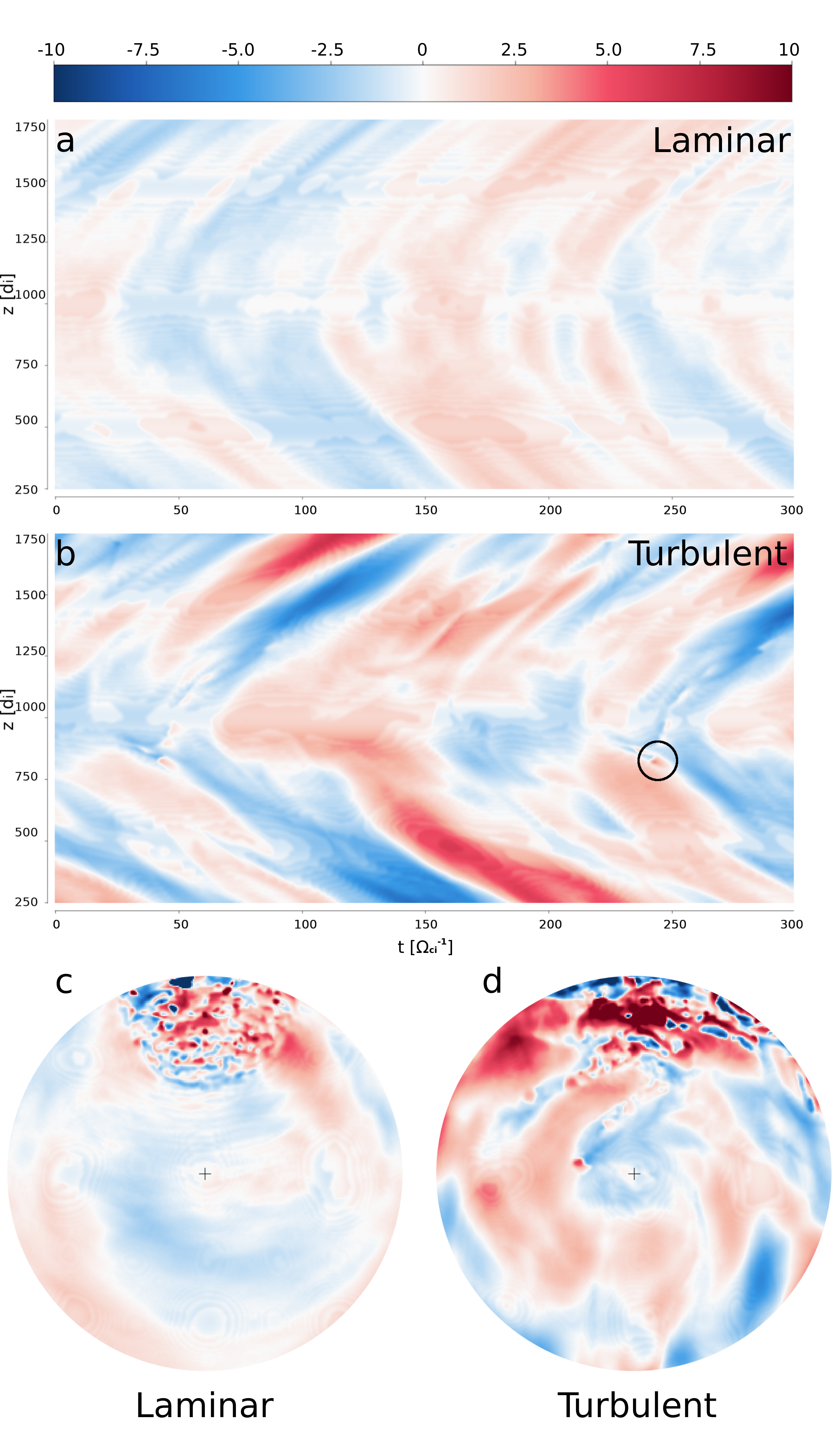}
\caption{Temporal evolution of the Bow shock position at each time step $t$ for the coordinates $(Y_\text{P}, Z)$ for the laminar (a) and turbulent (b) initial conditions. The corresponding planar projections (c,d) of the full 3D shock surface with respect to its time-averaged position at $t=250\,\Omega_\text{ci}^{-1}$. The colour code provides the deviation from the overall time-averaged position in $d_i$. The black circle points to the ``bubble'' appearing around $(1000, 800)$~$d_i$ in Figs. \ref{fig:comp_dens}e, \ref{fig:comp_B}e and \ref{fig:comp_bulkspeed}e and discussed in Fig.~\ref{fig:bubble_turbulent}. 
}
\label{fig:bound_pos}
\end{figure*}

\begin{figure*}
\centering
\includegraphics[width=\linewidth]{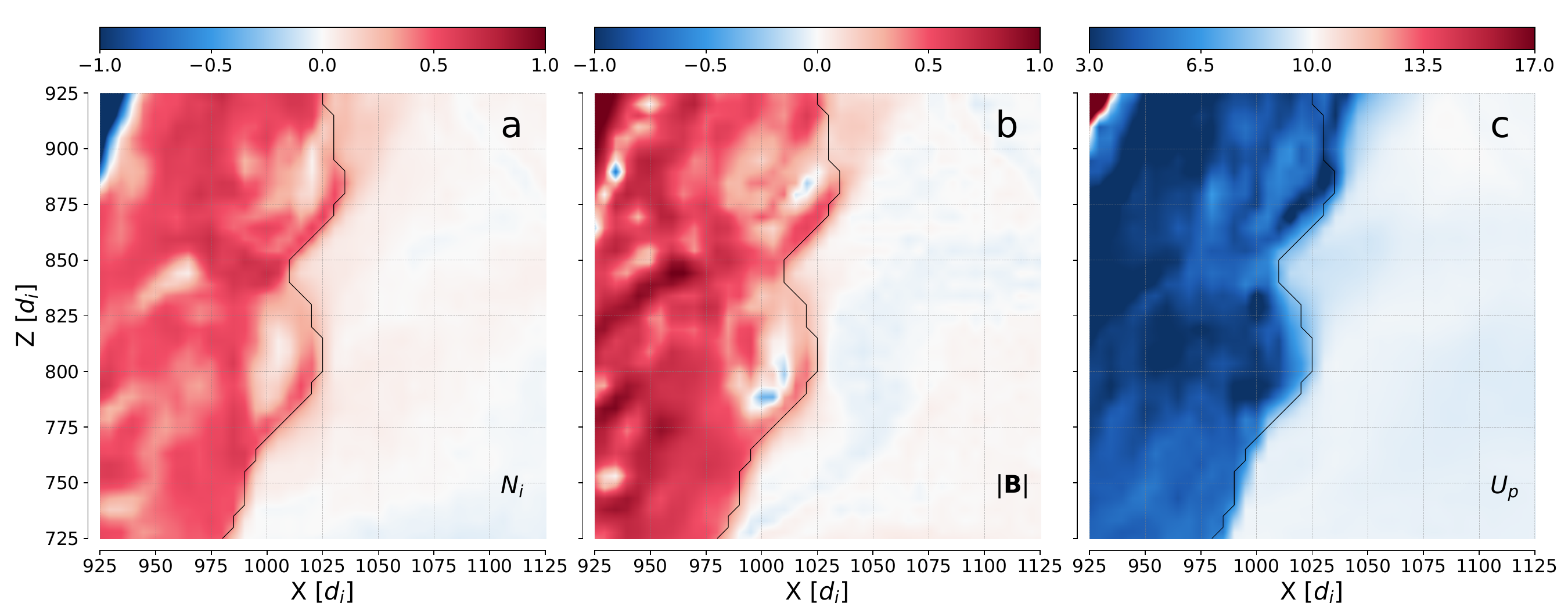}
\caption{Transient structure along the bow shock's location in the $X$--$Z$ plane of the turbulent simulation at $t=250\,\Omega_\text{ci}^{-1}$. (a): ion density $N_i$ (logarithmic scale, in units of $n_0$), (b): total magnetic field intensity $|\mathbf{B}|$ (logarithmic scale, in units of $B_0$) and (c) proton bulk speed $U_p$ (linear scale, in units of $C_A$, in the shock's reference frame).}
\label{fig:bubble_turbulent}
\end{figure*}

As already described, the proxy of the bow shock position introduced in the previous section is computed during runtime and at high cadence. This enables high-resolution analysis of the bow shock shape and evolution forced by the solar wind dynamics. This analysis is shown in Fig.\,\ref{fig:bound_pos}. To reduce the dimension of the problem, we first consider in Fig.\,\ref{fig:bound_pos}a, b the cut $Y=Y_\text{P}$ of the bow shock surface. The time evolution of this 1D cut along $z$ of the bow shock position is shown for both the laminar (Fig.\,\ref{fig:bound_pos}a) and turbulent (Fig.\,\ref{fig:bound_pos}b) solar wind dynamics. 
The bow shock position along the Sun-planet direction~$X$ with respect to its mean value over time, i.e., with respect to its time-averaged position, is shown in colour, with red (resp. blue) tones highlighting the times and positions at which the shock position is upstream (resp. downstream) of its average position. Using the same range of colours for the turbulent and laminar cases (the deviations from the average position are displayed between $-10\,d_i$ and $+10\,d_i$), we appreciate how the turbulence of the impinging solar wind induces much larger amplitude oscillations of the shock's surface. The variance of the values shown in Fig.\,\ref{fig:bound_pos}a is $0.8\,d_i$, while a greater variance of $2.0\,d_i$ is found in the turbulent case.

Second, we illustrate the deformation of the bow shock in Fig.\,\ref{fig:bound_pos}c, d by showing the planar projections of the full 3D shock surface position with respect to its time-averaged position at a given time $t=250\,\Omega_\text{ci}^{-1}$, for both laminar (Fig.\,\ref{fig:bound_pos}c) and turbulent (Fig.\,\ref{fig:bound_pos}d) solar wind dynamics. 
This representation of the bow shock deformation is similar to that of the position of a vibrating tambourine skin under the drumming action of the impinging solar wind. We observe local and global oscillations of the shock position. 
In the laminar solar wind dynamics case, the bow shock deformation is first observed at the bow shock nose and subsequently propagates from the nose towards the flanks, along the surface of the shock, creating the ``butterfly-shaped'' propagating structures in the $Z$--$t$ space observed in the top panels. 
In contrast, for the turbulent solar wind dynamics case, the bow shock deformation originates from multiple regions (not only the nose) depending on the solar wind dynamics and the local conditions at the shock. These deformations later propagate towards the flanks along the surface of the shock, generating an even more complex deformation pattern. This dynamics is reminiscent of the ubiquitous rippling observations at the Earth's quasi-parallel \citep{pollock2022}, quasi-perpendicular \citep{moullard2006} and oblique \citep{gingell2017} bow shock.

The quasi-parallel shock region exhibits significant variability, regardless of the initial solar wind conditions imposed in the simulations; the oblique and quasi-perpendicular shock surface variations are strongly enhanced in the turbulent solar wind case, with amplitudes reaching $\pm 10 d_i$, whereas in the laminar case, maximum amplitudes are much weaker ($\pm 2 d_i$). Beyond this much greater motion of the shock surface, turbulence is also responsible for the peculiar dynamics observed in confined regions of the shock front. At $t\sim 250\,\Omega_\text{ci}^{-1}$, close to the nose of the shock around $Z\sim 800\,d_i$, a small ``spot'' (circled in black in Fig.\,\ref{fig:bound_pos}b) departing from the average shock position is seen on the time series for the turbulent simulation. 
This transient structure is located at the bow shock interface around coordinates $(1000, 800)$~$d_i$ in panel e, corresponding to the $X$--$Z$ plane in Figs. \ref{fig:comp_dens}, \ref{fig:comp_B} and \ref{fig:comp_bulkspeed} and propagates along the shock's surface and inside the magnetosheath (Fig.\,\ref{fig:bound_pos}b). We show a zoom-in plot of this highly dynamic structure in Fig.\,\ref{fig:bubble_turbulent}, with density, magnetic field and bulk speeds, corresponding to a snapshot of the simulation when the structure has fully formed. It appears as a localised ``bubble'' of high-magnetic field, high-density plasma enclosing a much lower magnetic field and lower-density plasma. This suggests that, locally, a bubble of shocked solar wind plasma can impulsively penetrate inside the magnetosheath and start interacting with the local plasma there. 
Complexifying this picture, Fig.\,\ref{fig:bubble_turbulent}c also shows that the bulk speed inside this bubble is as low as its immediate surroundings with plasma already decelerated to magnetosheath-like speeds, whereas its density and magnetic field amplitude are closer to solar wind values. 
Although the analysis of this precise structure and others found in the quasi-perpendicular shock of the turbulent simulation is out of the scope of this study, it is interesting to notice that such signatures, characteristic of rippling and reformation processes which are usually found in quasi-parallel shocks, have also been seen in local hybrid simulations of plasma turbulence interacting with quasi-perpendicular shocks \citep{trotta2022transmission}.

%==============================
\subsection{Magnetosheath structure}

As can be seen in Fig.~\ref{fig:comp_dens} (plasma density), the thickness of the magnetosheath downstream of the quasi-parallel shock is smaller than downstream of the quasi-perpendicular shock for both laminar and turbulent solar wind conditions. 
This is consistent with THEMIS observations over a 5-year period \citep{Dimmock2013}, which uncovered an asymmetry in the Earth's magnetosheath between the dawn and dusk regions due to the nominal Parker spiral geometry. When comparing the laminar and turbulent runs, this asymmetry persists. 

In Figs.\,\ref{fig:comp_dens}a,c and \ref{fig:comp_B}a,c for the laminar run (left column), the magnetosheath region also exhibits large fluctuations in density and magnetic field magnitude, which fill all of the magnetosheaths, as expected from observations \citep{Narita2021}. Fluctuations are more coherent in the quasi-perpendicular region compared with the quasi-parallel region, where this coherence is mostly lost, and fluctuations become larger in size and amplitude. In the turbulent case, some of that coherence is further lost, as can be observed when comparing both columns in Figs. \,\ref{fig:comp_dens} and \ref{fig:comp_B}. This is very similar to what has been seen in numerical simulations of solar wind-comet interactions when considering the turbulent nature of the solar wind \citep{behar2023aa}.

The additional loss of coherence in fluctuations in the quasi-perpendicular magnetosheath due to solar wind turbulence may be explained by the transmission of large-scale solar wind turbulence structures across the shock. These structures are observed clearly in some regions immediately downstream of the bow shock, e.g., around coordinates ($400,~1100$)~$d_i$ and ($1000,~400$)~$d_i$ in the $Y$--$Z$ plane in Figs.~\ref{fig:comp_dens}f, \ref{fig:comp_B}f and \ref{fig:comp_bulkspeed}f. 
However, another possible explanation for the observed structures downstream of the quasi-perpendicular shock is the interaction of self-generated transients at the quasi-perpendicular shock, such as the high-density high-$B$-field ``bubble'' previously mentioned in Sect.~\ref{dynamics} around $(X,Z) \approx ($1000,~ 800$)$~$d_i$ in Figs.\,\ref{fig:comp_dens}e,~\ref{fig:comp_B}e~and~\ref{fig:comp_bulkspeed}e. 

For the quasi-parallel region, the presence of upstream solar wind turbulence increases the size and magnitude of the fluctuations in the magnetosheath, as can be seen for $Y\gtrsim 1300\,d_i$ in panel d, and for $Y\sim 1400\,d_i$ and $700\lesssim Z\lesssim 1300\,d_i$ in panel f of Figs.~\ref{fig:comp_dens}~and~\ref{fig:comp_B}. 
In this geometry, the fluctuations occurring on the downstream and upstream sides of the shock were already present in the laminar case, albeit in a less developed and intense manner (Figs.\,\ref{fig:comp_dens}c and \ref{fig:comp_B}c). Disentangling the effects due solely to solar wind dynamics turbulence from those inherited from the basic laminar conditions will require a dedicated analysis, which is outside the scope of the present study. These aspects will be explored in future research.

Figure\,\ref{fig:comp_bulkspeed} shows how the plasma in the wake of the bow shock is slowed down to speeds significantly below the upstream solar wind bulk speed of $U_\text{sw} = 10\,C_A$, with white marking the reference solar wind speed. In the quasi-perpendicular side of the magnetosheath in the laminar case (Fig.\,\ref{fig:comp_bulkspeed}a), fluctuations in plasma speed provide a ``baseline'' level of the magnetosheath natural turbulence, with striations appearing perpendicular to the shock surface in the immediate wake of the shock front (as clearly seen in Fig\,\ref{fig:comp_bulkspeed}c). In contrast, bulk plasma speed fluctuations are much increased for the turbulent case as compared to the laminar case, with large wavy structures developing almost parallel to the shock surface behind the terminator line (Fig.\,\ref{fig:comp_bulkspeed}d) and superimposed to the natural turbulence of the magnetosheath (Fig.\,\ref{fig:comp_bulkspeed}f). Deeper in the quasi-perpendicular magnetosheath, the plasma is further compressed and witnesses increased speed near the modelled magnetopause. In general, downstream of the quasi-perpendicular shock, the plasma velocity fluctuates much more in the turbulent case than in the laminar case, as especially seen in the flanks (Fig.\,\ref{fig:comp_bulkspeed}f), with large defined structures possibly modulated by the global scale turbulence upstream of the shock.

In the quasi-parallel side of the magnetosheath, the conclusions already drawn for Figs.\,\ref{fig:comp_dens} and \ref{fig:comp_B} hold: fluctuations in the ion foreshock region increase substantially, with a loss of coherence of the backstreaming ions that create the characteristic Ultra Low Frequency (ULF) waves populating the foreshock. Streams of low bulk speeds (in dark blue, Fig.\,\ref{fig:comp_bulkspeed}d) appear upstream of the shock, corresponding to relatively low magnetic field intensities-low plasma densities. %Downstream of the quasi-perpendicular shock, the plasma velocity fluctuates much more in the turbulent case than in the laminar case, as especially seen in the flanks (Fig.\,\ref{fig:comp_bulkspeed}f), with large defined structures possibly modulated by the global scale turbulence upstream of the shock.

%--- transition to elegantly go back upstream, specifically to the ion foreshock
While the solar wind mainly crosses the quasi-perpendicular part of the bow shock, some of it is actually reflected in the quasi-parallel part, forming the so-called ion foreshock region upstream of the quasi-parallel shock region. We now focus on this region.  

%==============================
\subsection{Dynamics within the ion foreshock}
\label{sec:ion_foreshock}

In this section, we discuss the influence of the turbulent nature of the solar wind on ion foreshock.
Because the bow shock is supercritical and collisionless, solar wind ions are expected to be reflected in the bow shock region quasi-parallel to the solar-wind magnetic field. This is well modelled in our kinetic hybrid simulations, and the resulting ion foreshock is observed upstream of the shock, for $Y>1400$~$d_i$ in panels a,d in Figs.~\ref{fig:comp_dens}, \ref{fig:comp_B} and~\ref{fig:comp_bulkspeed}, as expected.

\begin{figure*}
    \centering
    \includegraphics[width=0.9\linewidth]{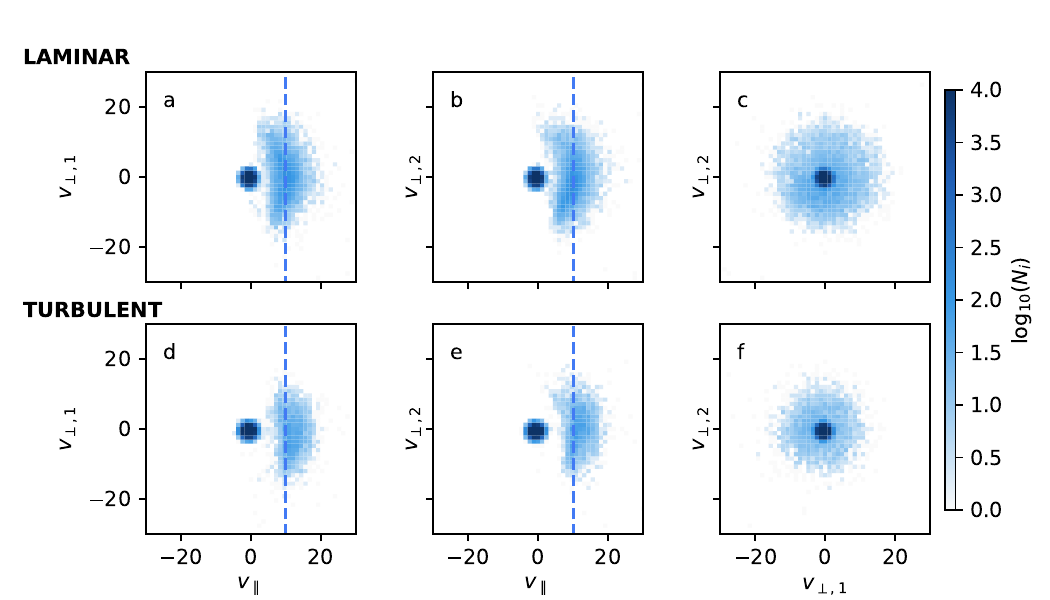}
    \caption{Ion velocity distribution function (VDF) for the laminar (panels a-c) and turbulent case (panels d-f) in logarithmic scale. The VDF is plotted in a reference frame aligned with the local magnetic field and moving with the solar wind. The plots show the number of macroparticles integrated in the out-of-plane direction. The blue dashed line corresponds to the opposite of the solar wind speed in the planet reference frame.}
    \label{fig:vdf}
\end{figure*}

The solar wind ions reflected by the bow shock are seen in the ion velocity distribution functions (VDFs) shown in Fig.~\ref{fig:vdf}, both for the laminar and turbulent solar wind. 
The VDF is computed in a cubic box centred in ${\bf r}_{0,vdf}(x,y,z)=(860,1620,1020)d_i$ and having size $40 d_i\times40 d_i\times40 d_i$. The box size is chosen to ensure enough statistics on the particle beam. Figure~\ref{fig:vdf} displays the VDFs in a reference frame oriented as  
$\hat{\bf{e}}_{\parallel} = \bf{B}/{\bf{|B|}}$, $\hat{\bf{e}}_{\perp,1} =-\bf{V}\times\bf{B}/(|\bf{V}\times\bf{B}|)$, and $\hat{\bf{e}}_{\perp,2}=\hat{\bf{e}}_{\parallel}\times\hat{\bf{e}}_{\perp_1}$, where $\bf{B}$ and $\bf{V}$ are the local magnetic and ion velocity fields, i.e. their box-averaged value in the box where the VDF is computed. In both laminar and turbulent solar wind conditions, we observe the characteristic solar wind core population centred at the origin of the coordinates and a less populated beam moving on average in the $+{\hat{\bf{e}}}_{\parallel}$ direction, as indicated by the dashed blue line. %moving on average in the ${\hat{\bf{e}}}_{\parallel}$ direction opposite to the solar wind speed in the object's reference frame, as indicated by the dashed blue line. 
The beam velocity is $\varv_\text{beam}= - U_\text{sw} = 10$ in Alfvén speed (code) units. The beam width is comparable in the two perpendicular directions and the VDF is thus close to gyrotropic (Fig.\,\ref{fig:vdf}c,f). The two VDFs (in laminar and turbulent solar wind conditions) look quite similar in the selected location. However, the particle density in the beam corresponding to reflected particles is reduced in the turbulent case compared to the laminar one. We observed such a feature everywhere in the ion foreshock.

\begin{figure*}
    \centering
    \includegraphics[width=0.9\linewidth]{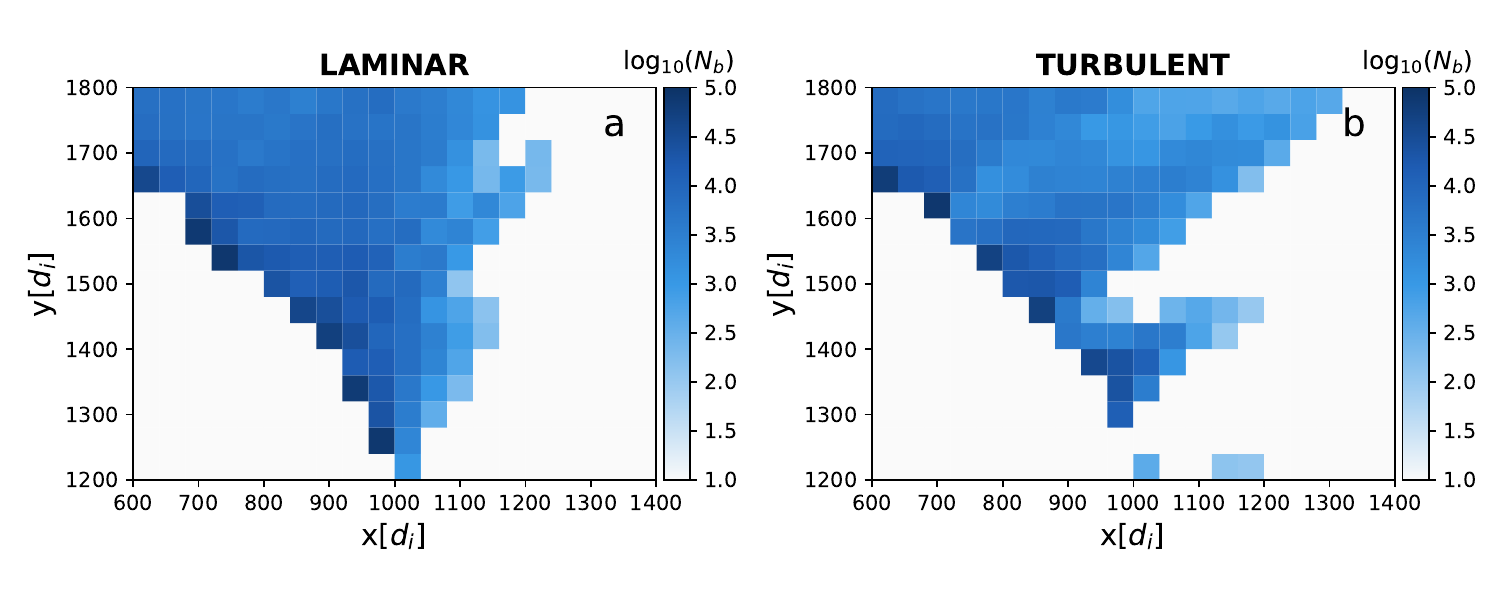}
    \caption{Ion beam density $N_b$ in the foreshock region for a) laminar and b) turbulent solar wind dynamics conditions. The colour bar is in logarithmic scale; $N_b$<10 in white regions. 
    }
    \label{fig:comp_nb}
\end{figure*}

The presence of solar wind turbulence influences the spatial distribution of the reflected beam itself. Figure~\ref{fig:comp_nb} shows its density $N_b$ in the foreshock region for the two simulations. The plot is obtained using the following procedure. The VDF is computed in boxes of $40 d_i \times 40 d_i \times 40 d_i$, forming a grid in the physical space. 
For each set of particles located inside one box, particles having a speed $\varv>5 C_A$, where $C_A$ is the Alfvén speed in the pristine solar wind in our simulation, are used to compute the beam density. We observe that the fluctuations in the reflected beam density are much more pronounced in the turbulent (Fig.\,\ref{fig:comp_nb}b) than in the laminar (Fig.\,\ref{fig:comp_nb}a) case. We argue that this behaviour is due to two factors: in the turbulent case as compared with the laminar, i) the foreshock base region is more inhomogeneous (as shown in Fig.~\ref{fig:B_topology_qpara}), and ii) the magnetic field line diffusion in the direction perpendicular to the mean magnetic field is enhanced. The combination of these two processes influences the transport of particle beams, resulting in a loss of coherence of the beam itself when moving away from the bow shock and a more patchy density distribution in the turbulent case. 

To elucidate the global picture of this process, Fig.~\ref{fig:B_topology_qpara} shows a 3D rendering of magnetic field lines in the foreshock region. We observe that the laminar case's magnetic field lines appear to align with the direction of the solar-wind magnetic field. The field lines oscillate due to beam-induced waves in the foreshock, typical of the right-hand polarized ULF waves seen at Earth and arising from wave-particle interactions \citep{narita2004}. In contrast, the field line topology appears much more complex in the turbulent case. The oscillations in the laminar case are observed only close to the shock base and disappear while moving away from it towards the solar wind. Moreover, in the turbulent case, some magnetic field lines crossing the foreshock region have footprints outside the quasi-parallel shock base. This implies that part of the plasma in the foreshock region comes from outside the foreshock where particle reflection has been less efficient. This complex magnetic topology results in the patchy distribution of the ion beam density in the foreshock region, as reported in Fig.\,\ref{fig:comp_nb}b. 

\section{Conclusion}\label{conclusion}

What is the impact of the turbulent nature of the solar wind on its interaction with a magnetized planet like the Earth or Mercury? How do the dynamics of solar wind turbulence affect the bow shock location and shape? 
How is the solar wind turbulence itself modified by the bow shock crossing? 
What is the influence of the solar wind turbulence on the ion foreshock?
%--- Here state the novelty of the study
To address these questions, we performed the first 3D global simulation of the interaction between a turbulent solar wind and a magnetized planet. We investigated the influence of turbulence on the bow shock shape and dynamics, the structure of the magnetosheath, and the ion foreshock. 

Regarding the bow-shock dynamics, larger fluctuations in the shock's position are observed as compared to the laminar case. Additionally, we have shown that while in the laminar case the deformation of the bow shock outside of the quasi-parallel region originates solely at the nose of the shock, in the turbulent case deformations are triggered in multiple regions depending on solar wind dynamics and local conditions. These deformations propagate along the shock's surface towards the flanks, resulting in a more complex pattern. Consequently, the oscillations in the surfaces of oblique and quasi-perpendicular shocks are significantly amplified in turbulent solar wind conditions. Our study also shows that bubble-like plasma structures can form in the quasi-perpendicular shock area, where they start interacting with the local plasma in the magnetosheath. Further investigation into this phenomenon is deferred to future research. 

The magnetosheath structure under laminar and turbulent solar wind conditions exhibits similar behaviour with a spatial asymmetry between the quasi-parallel and quasi-perpendicular sides of the shock in agreement with observational statistical studies \citep{Dimmock2013}. The main effect of the turbulent solar wind dynamics on the magnetosheath is, on average, to diminish the coherence of the $B$-field and density fluctuations and enhance their amplitude, which is qualitatively consistent with the observed transmission through the bow shock of the turbulence inherited from the solar wind. In the ion bulk speeds, we also observed the appearance of structures almost parallel to the shock surface and superimposed to the perpendicular structures containing relatively larger ion speeds that populate the laminar case simulation.
In the future, we aim at exploring in more detail how 3D solar wind structures are processed by the shock \citep[following, e.g.,][]{trotta2022transmission} and investigate whether the relaxed equilibrium states, typical of the turbulent phenomenology, that are observed in the magnetosheath are locally generated or may originate from the solar wind \citep{pecora2023relaxation}.

In the ion foreshock region, the presence of upstream turbulence influences the spatial properties of the reflected ion beam. Specifically, this ion beam in the turbulent case is more inhomogeneously distributed in space and extends less further upstream from the shock than in the laminar one due to the enhanced complexity of the magnetic field lines. Furthermore, we have shown that turbulence and beam-induced fluctuations in the foreshock region may exist for the solar wind turbulence level considered in our simulation. We may expect their presence and importance in the foreshock region to vary with the amplitude of the turbulence advected by the solar wind. A systematic study of the interplay between the two will require more simulations where the amplitude of the upstream solar wind turbulence is varied; this is also left for future work.

Menura's distinctive approach, reproducing the global interaction of a turbulent solar wind with compact objects, including planetary magnetospheres,  induced or not, marks a significant advancement in our theoretical description of the near-Earth environment. Multi-satellite missions such as ESA/Cluster, NASA/Time History of Events and Macroscale Interactions during Substorms (THEMIS) or, more recently, the NASA/Magnetospheric Multiscale Mission (MMS), all continue to investigate with increasing temporal and spatial resolution the Earth's magnetosphere, magnetosheath and near-Earth solar wind and may benefit from numerical work such as that we presented here. 

Further, Menura is a parallel code based on and running on GPUs. We expect that, in the future, due to increased computational power, simulation at the full Earth scale may become feasible. Meanwhile, planetary magnetospheres of smaller sizes may be studied with full-scale simulations. These upcoming simulations are timely, considering that the ESA-JAXA/BepiColombo mission will reach Mercury in December 2025.  
Additionally, due to the ease with which different temporally variable initial conditions can be imposed in Menura, future studies could also include modelling magnetic clouds (coming from Coronal Mass Ejections) and stream interaction regions. 

Finally, numerical studies of global dynamics that consider kinetic effects, such as presented in this work, may also be relevant to many other astrophysical systems constituted by a compact object, with or without an intrinsic magnetic field, interacting with a non-laminar flow.

\begin{acknowledgements}
      
    E. Behar acknowledges support from the Swedish National Research Council, Grant 2019-06289. This work was granted access to the HPC resources of IDRIS under the allocation 2021-AP010412309 made by GENCI. 
    P.H. acknowledges support from CNES APR. This work was granted access to the HPC resources of IDRIS under the allocation 2024-AD010412990R1 made by GENCI. 
    C. Simon Wedlund is funded by the Austrian Science Fund (FWF) 10.55776/P35954.  
    L. Preisser was supported by the Austrian Science Fund (FWF): P 33285-N.
    %G. Ballerini acknowledges the support from the Italian supercomputing center CINECA through the ISCRA grant.
    F.P. acknowledges support from the National Research Council (CNR) Short Term Mobility (STM) 2023 program and from the Research Foundation – Flanders (FWO) Junior research project on fundamental research G020224N.
    We acknowledge the CINECA award under the ISCRA initiative, for the availability of high performance computing resources and support.
\end{acknowledgements}

\bibliographystyle{aa}
\bibliography{eb_biblio} 

\end{document}